# Certainty of outlier and boundary points processing in data mining

Elyas Rashno[1], Sanaz Saki Norouzi[1], Behrouz Minaei-bidgoli[1] and Yanhui Guo[2]

[1] Department of Computer Engineering, Iran University of Science and Technology, Tehran, Iran

E-mail: elyas.rashno@gmail.com

[2] Department of Computer Engineering, University of Illinois Springfield, Springfield, USA

*Abstract*— Data certainty is one of the issues in the real-world applications which is caused by unwanted noise in data. Recently, more attentions have been paid to overcome this problem. We proposed a new method based on neutrosophic set (NS) theory to detect boundary and outlier points as challenging points in clustering methods. Generally, firstly, a certainty value is assigned to data points based on the proposed definition in NS. Then, certainty set is presented for the proposed cost function in NS domain by considering a set of main clusters and noise cluster. After that, the proposed cost function is minimized by gradient descent method. Data points are clustered based on their membership degrees. Outlier points are assigned to noise cluster and boundary points are assigned to main clusters with almost same membership degrees. To show the effectiveness of the proposed method, two types of datasets including 3 datasets in Scatter type and 4 datasets in UCI type are used. Results demonstrate that the proposed cost function handles boundary and outlier points with more accurate membership degrees and outperforms existing state of the art clustering methods.

## I. INTRODUCTION

Data certainty is one of the issues that we face in the real world due to the deviation in the original data. For example, in sensor networks [1] and services which are based on locations [2] such as object tracking, it is almost impossible to have the exact location of the object at all times, and it is just an example of certainty so different source of certainty must be considered to have an accurate result [3]. Data certainty can be categorized to *existential certainty* and *value certainty*, first type expressed certainty by existence or nonexistence of the object or data tuple (the existence or not (absence) an object or data is certain). For example, in a relational database, a tuple has a probability which is show the confidence of its presence [4, 5]. In the second type of certainty, a data is limited to the values around it, and then can be modelled by using the probability density function of its values [1, 6]. According to the above content, data mining can be considered in two groups: mining on certain data, and mining on uncertain data. The second one also can be categorized to some main categories such as association rule mining, data classification, data clustering and other data mining methods [3]. Clustering is a compact representation of data with a limited number of clusters which has been applied in many applications including pattern recognition, computer vision and image processing, taxonomy, geology, business and medicine [7]. Generally, data clustering is divided into crisp and fuzzy methods referred as hard and soft clustering, respectively. In fuzzy-based methods, each data point is assigned to all clusters with different membership degrees while in crisp methods it is enforced to be assigned to only one cluster. K-means and fuzzy c-means are the most popular crisp and fuzzy clustering methods, respectively [8]. Little research has been reported in data clustering on uncertain data. In [9], EM algorithm is used for addressing fitting mixture densities. In [10] a new method which is named as possibilistic c-means (PCM) proposed to make changes on FCM by making the membership small if the point is far from the cluster center by adding a penalty term. In [11], to make the PCM algorithm flexible to set initial cluster number, the APCM is proposed. In [12], a new method based on PCM (NPCM and UPCM) is proposed to handled uncertainty in clustering by making changes in parameters. In [13], a new dynamic technique for hierarchical clustering proposed for data uncertainty. Neutrosophy (NS) is a branch of philosophy which studies the nature and scope of the neutralities and their interactions. The basis of neutrosophic logic and neutrosophic set was presented in the first world neutrosophic publications in [14], [15], [16]. In the recent years, neutrosophic has been applied in different image processing applications such as image segmentation [17-20], CBIR [21-22], edge detection [23], image thresholding [24], image clustering [25], image denoising [26], retinal image analysis [27-31] and [39-40], and breast image analysis [32].

The main contribution of this work is to model certainty in neutrosophic set domain for clustering application. This model computes certainty for each data point and let us interpret it as either outlier, boundary or normal data point and then decide how to assign it to clusters. Note that these types of data points are the main challenge of all clustering methods. In this research, a new definition of certainty is presented for a proposed cost function with two types of clusters including main clusters for normal and boundary points and noise cluster for outlier points. Data certainty is presented for a proposed cost function in NS domain. The proposed cost function assigns data points with low certainty to noise cluster and other data points to main clusters. It also assigns the same membership degrees to main clusters for boundary points. The proposed cost function is minimized with gradient descent method which leads to data clustering. This model can be used in any type of data including, scatter data, image, speech, and datasets with a set of features and class labels. The rest of this paper is organized as follows: Section II and III present the neutrosophic set and proposed method, respectively. Results are discussed in Section IV. Discussion is presented in Section V. Finally, this work is concluded in Section VI.

## II. NEUTROSOPHIC SET

NS is a powerful framework of neutrosophy in which neutrosophic operations are defined from a technical point of view. In fact, for each application, neutrosophic sets are defined as well as neutrosophic operations corresponding to that application. Generally, in neutrosophic set $A$, each member $x$ in $A$ is denoted by three real subsets true, false and indeterminacy in interval [0, 1] referred as $T$, $F$ and $I$, respectively. Each element is expressed as $x(t, i, f)$ which means that it is $t\%$ true, $i\%$ indeterminacy, and $f\%$ false. In each application, domain experts propose the concept behind true, false and indeterminacy. To use NS in image processing domain, the image should be transferred into the neutrosophic set domain. Although the original method for this transformation was presented by Guo et all. [17], these methods completely depend on the special application in image processing. An image $g$ has a dimension of $M \times N$. $g$ can be shown with three subsets: $T$, $I$ and $F$ in NS domain. So, pixel $p(i,j)$ in $g$ is shown with $PNS(i, j) = \{T(i, j), I(i, j), F(i, j)\}$ or $PNS (t, i, f)$. $T$, $I$ and $F$ indicate white, noise and black pixel sets, respectively. $PNS (t, i, f)$ provides useful information about white, noisy and black percentage in this pixel that is $t\%$ to be a white pixel, $i\%$ to be a noisy pixel and $f\%$ to be a black pixel. $T$, $I$ and $F$ are computed as follows [2-3].

$$T(i,j) = \frac{\overline{g(i,j)} - \overline{g_{min}}}{\overline{g_{max}} - \overline{g_{min}}} \quad (1)$$

$$F(i,j) = 1 - T(i,j) \quad (2)$$

$$I(i,j) = \frac{\delta(i,j) - \delta_{min}}{\delta_{max} - \delta_{min}} \quad (3)$$

$$\overline{g(i,j)} = \frac{1}{w^2} \sum_{m=-\frac{w}{2}}^{m=\frac{w}{2}} \sum_{n=-\frac{w}{2}}^{n=\frac{w}{2}} g(i+m, j+n) \quad (4)$$

$$\delta(i,j) = |g(i,j) - \overline{g(i,j)}| \quad (5)$$

where $g$ is gray scale image, $\overline{g}$ is filtered image g with average filter, $w$ is window size for average filter, $\overline{g_{max}}$ and $\overline{g_{min}}$ are the maximum and minimum of the $\overline{g}$, respectively, $\delta$ is the absolute difference between $g$ and $\overline{g}$, $\delta_{max}$ and $\delta_{min}$ are the maximum and minimum values of $\delta$, respectively.

## III. PROPOSED METHOD

In this work, two sets of clusters including main and noise clusters are considered for data points. Outlier points are assigned to noise cluster and boundary points are assigned to main clusters with almost same membership degrees. A new cost function for data clustering in neutrosophic domain is proposed in (6):

$$N(T,F,C) = \sum_{i=1}^{n}\sum_{j=1}^{k}(T_{i,j}(1-D_i))\|X_i - C_j\|^2 + \sum_{j=1}^{k}(F_i D_i)(K - \sum_{j=1}^{K}\|X_i - C_j\|^2) \quad (6)$$

where $T_{i,j}$ and $F_i$ are the membership degrees of data point $i$ to main cluster $j$ and noise cluster, respectively. All membership degrees are considered in interval $0 < T_{i,j}$ & $F_i < 1$ in neutrosophic domain with constraint $\sum_{j=1}^{K} T_{i,j} + F_i = 1$.

$K$ is the number of main clusters. Finally, $D_i$ is the certainty of data point $i$ which is defined by considering the distance of $i$ from its neighbors in (7)-(9).

$$D(i) = \begin{cases} 1 - \frac{InCircle(i)}{N/K} & \text{If } InCircle(i) < T_r \\ \alpha & \text{Otherwise} \end{cases} \quad (7)$$

$$InCircle(i) = \sum \delta(d(i,j), Eps) \quad (8)$$

$$\delta(a,b) = \begin{cases} 1 & a \leq b \\ 0 & a > b \end{cases} \quad (9)$$

where $N$ is the size of dataset, $T_r$ is a constant number as a threshold value, $Eps$ is a distance threshold and $d(i,j)$ is the Euclidean distance between data point $i$ and $j$. This definition for certainty of data point $i$ can be interpreted as: the more number of neighbors of data point $i$ in a circle around $i$, the bigger certainty is achieved. It is clear that this idea assigns certainty near to 0 for outlier pixels and near to 1 for points inside the main clusters. The main idea behind the proposed cost function is that, minimizing this function leads data point $i$ to be assigned to the main cluster $k$ if it satisfies two conditions including: point $i$ has the minimum distance from this cluster center and a big certainty $D_i$. Similarly, there are also two conditions for point $i$ to be assigned to noise cluster: (a) having the maximum sum distance from all main clusters $\sum_{j=1}^{k}\|X_i - C_j\|^2$ and (b) having a small certainty $D_i$. For minimizing the proposed cost function, the Lagrange function is constructed by (10):

$$N(T,F,C) = \sum_{i=1}^{n}\sum_{j=1}^{k}(T_{i,j}(1-D_i))\|X_i - C_j\|^2 + \sum_{j=1}^{k}(F_i D_i)(K - \sum_{j=1}^{k}\|X_i - C_j\|^2) - \sum_{i=1}^{n}\lambda_i(\sum_{j=1}^{k}T_{i,j} + F_i - 1) \quad (10)$$

For cost function minimization, gradient descent approach is used. Therefore:

$$\frac{\partial N}{\partial T_{i,j}} = ((1-D_i)T_{i,j})\|X_i - C_i\|^2 - \lambda_i \quad (11)$$

$$\frac{\partial N}{\partial F_i} = (D_i F_i)(K - \|X_i - C_i\|^2) - \lambda_i \quad (12)$$

$$\frac{\partial N}{\partial C_j} = -(D_i T_{i,j})(X_i - C_j) + (1-D_i)F_i(X_i - C_j) \quad (13)$$

By considering $\frac{\partial N}{\partial T_{i,j}} = 0$, $\frac{\partial N}{\partial F_i} = 0$ and $\frac{\partial N}{\partial C_j} = 0$ we can obtain equations for update $T_{ij}, F_i$ and $C_j$ as follows:

$$T_{i,j} = \frac{\lambda_i}{(1-D_i)\|X_i - C_j\|^2} \quad (14)$$

$$F_i = \frac{\lambda_i}{(D_i)(K - \|X_i - C_j\|^2)} \quad (15)$$

$$C_j = \frac{\left[\sum_{i=1}^n ((1-D_i)T_{i,j}) - \sum_{i=1}^n (D_i F_i)\right] X_i}{\left[\sum_{i=1}^n ((1-D_i)T_{i,j}) - \sum_{i=1}^n (D_i F_i)\right]} \quad (16)$$

The proposed clustering algorithm is summarized as follows:

**Algorithm 1**
1: Initialize membership degrees randomly and set $K$, $Eps$ and $Tr$.
2: Compute $D_i$ for for all data points.
3: Update $T_{ij}$, $F_i$ and $C_i$.
4: Check the stop condition, if $|N(T,F,C)^{(k)} - N(T,F,C)^{(k-1)}| < \varepsilon$ then stop, otherwise go to step 3.
5: Assign each data point into boundary cluster if the first two membership degrees $T_{ij}$ and $T_{ik}$ are between $t$ and $(1-t)$, otherwise assign it to a cluster which data point $i$ has the maximum membership degree to it.
6: end.

IV. EXPERIMENTAL SETUP AND RESULTS

The proposed clustering method evaluated on 3 local scatter datasets referred as X13, X37 and X43 and 4 datasets from UCI including "Ionosphere", "Haberman's survival", "Heart" and "Glass". Note that Parameters in the proposed method are set to quantities as follows: $T_r=4$, b=4 and $\alpha=0.95$.

*A. Scatter datasets*

The local scatter datasets visualize how the proposed method can handle outlier and boundary data points. In X13 dataset in Fig. 1, data points 1-8 are in the main clusters, 9 is a boundary point and 10-13 are outliers. Assigned membership degrees to each data point are demonstrated in Fig. 2. It is clear that the proposed method assigns points 1-4, 5-8 and 10-13 to $C_1$ (main cluster 1), $C_2$ (main cluster 2) and $F$ (noise cluster), respectively with a high level of certainty. Point 9 is assigned to $C_1$ and $C_2$ with almost same membership degrees which is interpreted as boundary point. Membership degrees for all data points are reported in Table 1. In X37, 3 clusters are considered with 6 boundary points (3 data points between $C_1$-$C_2$ and 3 data points between $C_2$-$C_3$) and 4 outliers (see Fig. 3). In this case, more boundary points and outliers are considered to evaluate the results of the proposed clustering method. As it is clear from membership degrees depicted in Fig. 4, all data points are assigned to correct clusters.

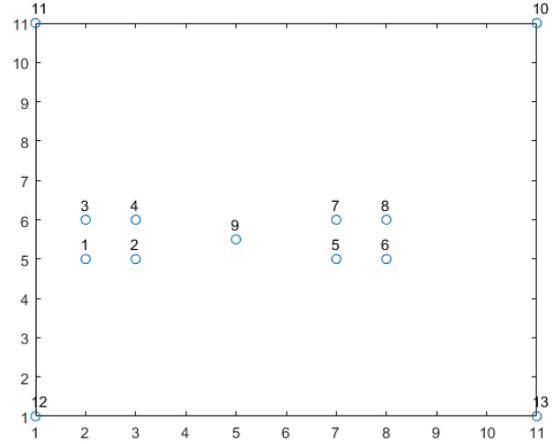

Figure 1.    X13 dataset

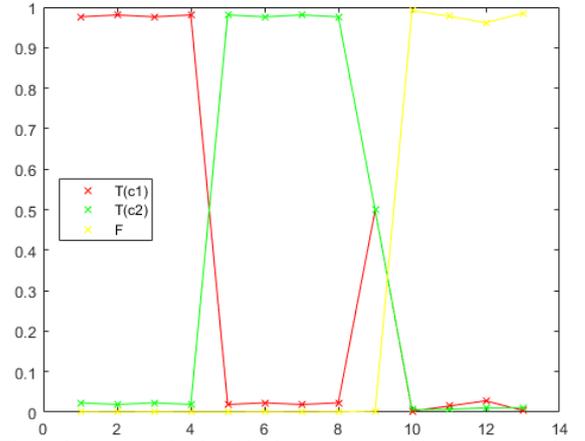

Figure 2.    Membership degree computed by the proposed method in X13.

Table 1.  X13 result

| Point N | Tc1 | Tc2 | F | Prop. |
|---|---|---|---|---|
| 1 | 0.9762 | 0.0231 | 0.0007 | C1 |
| 2 | 0.9808 | 0.0189 | 0.0004 | C1 |
| 3 | 0.9762 | 0.0231 | 0.0007 | C1 |
| 4 | 0.9808 | 0.0189 | 0.0004 | C1 |
| 5 | 0.0189 | 0.9808 | 0.0004 | C2 |
| 6 | 0.0231 | 0.9762 | 0.0007 | C2 |
| 7 | 0.0189 | 0.9808 | 0.0004 | C2 |
| 8 | 0.0231 | 0.9762 | 0.0007 | C2 |
| 9 | 0.4987 | 0.4987 | 0.0026 | **boundary** |
| 10 | 0.0026 | 0.0059 | 0.9916 | **outlier** |
| 11 | 0.0151 | 0.0071 | 0.9778 | **outlier** |
| 12 | 0.0279 | 0.0106 | 0.9614 | **outlier** |
| 13 | 0.0041 | 0.0109 | 0.985 | **outlier** |

Finally, in X43, the position of the main clusters and boundary points are changed which is shown in Fig. 5. Points

37 and 39 are boundary points between $C_1$-$C_2$ and $C_1$-$C_3$, respectively. Point 38 is a boundary point between $C_1$-$C_2$-$C_3$ which is more challenging for clustering. The membership degrees for all data points are depicted in Fig. 6. As it is clear from membership degrees assigned to boundary points, these data points are detected with lower certainty in comparison with X13 and X37.

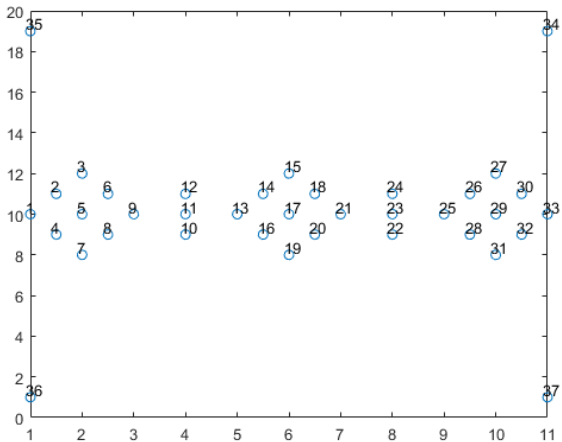

Figure 3.  X37 dataset

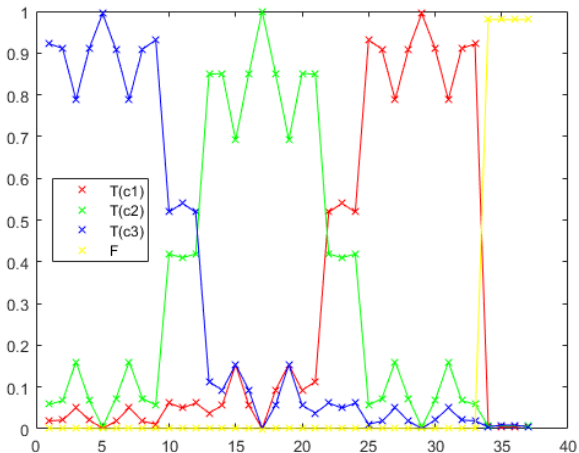

Figure 4.  Membership degree computed by proposed method in X37

*B. UCI datasets*

UCI datasets, as standard datasets in machine learning community, are used in this research to evaluate and compare the result of the proposed method with other methods applied in UCI datasets. In this research, "Ionosphere ", "Haberman's survival", "Heart" and "Glass" datasets are selected among other datasets in UCI. The Johns Hopkins Ionosphere; referred as Ionosphere dataset; is a binary classification dataset which contains 351 data points that 225 and 126 objects are for the first and the second data classes, respectively. This dataset is a 34 dimensional dataset but one attribute has a completely zero values, so it is discarded and values along each dimension were normalized to have Mean 0 and Standard deviation 1.

Haberman's Survival referred as Haberman dataset has been collected from a study on the survival of breast cancer patients that shows they survive or die in surgery. This dataset is a 3 dimensional dataset that includes 306 instances that 225 objects were assigned to the first class. The third dataset in UCI is Heart dataset which shows the presence of heart disease in the patient. This dataset contains 75 attributes but published experiments used 13 features of them. It includes 303 instances and all data points assigned to class 0 (no presence) to 4. The last dataset is Glass data frame with 214 observation containing examples of the chemical analysis of 7 different types of glass. The problem is to forecast the type of class on basis of the chemical analysis. The study of classification of types of glass was motivated by criminological investigation. At the scene of the crime, the glass left can be used as evidence. This dataset contains 214 9-dimensional samples in 6 clusters. Table 2 summaries number of features, number of classes, samples in each cluster and number of objects in each dataset. Accuracy of the proposed method, FCM [33], PFCM [34], Soft-DKM [35], WEFCM [36], CDFCM [37] and DPFCM [38] methods are summarized in Table 3. The proposed method outperforms other methods in "Ionosphere ", "Haberman's survival", "Heart" and "Glass" datasets with the accuracy of 79.12%, 80.04%, 83.34% and 59.43%, respectively.

V. DISCUSSION

The proposed scheme addresses two challenges in data clustering which are highly correlated to each other. First, there are points between boundary cluster and a main cluster. Since such points are located in the same distance from the center of the main cluster and the center of boundary cluster, they are not assigned to the main cluster with a high certainty. Second, such points displace center of the main clusters because the proposed equation for cluster center calculation considers all data points with a membership for cluster center computation. In the proposed method, these issues are addressed by proposing a cost function in which noise cluster is considered and boundary cluster is ignored. Reported results in Scatter datasets demonstrate how the proposed method handles boundary points and points between boundary points and main cluster centers. Boundary points have a same distance from main cluster centers. In clustering algorithms, when cluster centers are converged; without significant changes in subsequent iterations; for data point $x_i$, term $\|x_i - C_j\|$ has the same quantity for all cluster centers ($j = 1,2,…,K$). It leads membership degree $x_i$ to cluster center $j$ ($T_{ij}$) to be almost same for all clusters. We have used this property to distinguish boundary points.

The main problem in clustering is determining the number of main clusters ($K$). This issue is domain-specific and should be determined under expert supervision. In each dataset among scatter and UCI datasets, data are grouped into specific number of clusters ($K$). X13: K=2, X37: K=3, Ionosphere, Haberman and Heart: K=2 and Glass: K=6. Here, in each experiment $K$ is determined from pre-defined context knowledge. It should be noted that inappropriate $K$ affects clustering results significantly. Also, in the proposed method, the more number of clusters in datasets, the more failures appear in boundary and noisy data points modeling in NS domain. In Glass datasets with 6 clusters, the proposed method clusters data with the accuracy of 59.43%. The reason is that in this dataset, the distance of boundary data points should be considered from 6 clusters simultaneously while in Heart dataset this constraint is for 2 clusters.

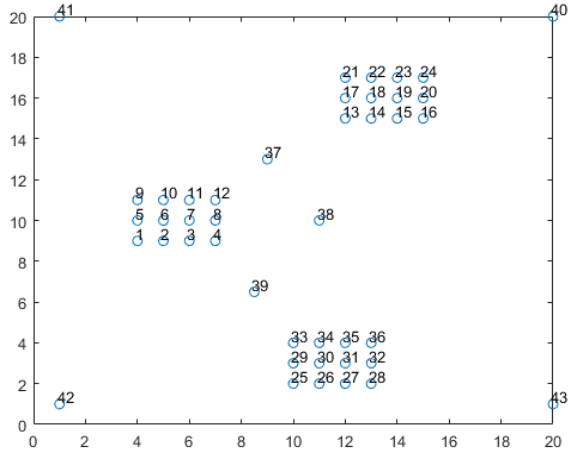

Figure 5. X43 dataset

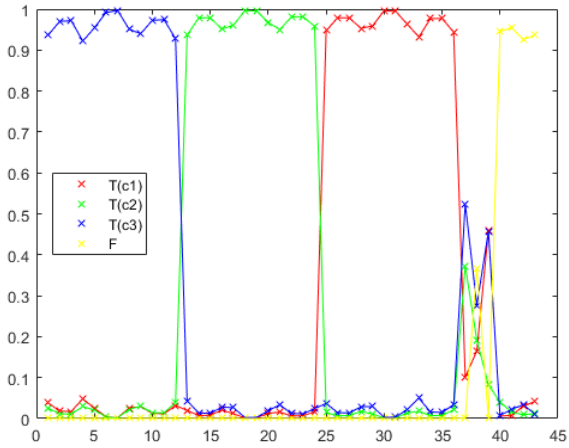

Figure 6. Membership degree computed by the proposed method

Table 2. Summary of datasets characteristic

| Dataset | No. of Features | No. of Clusters | Samples in each Cluster | No. of objects |
|---------|-----------------|-----------------|-------------------------|----------------|
| Ionosphere | 34 | 2 | 126,225 | 351 |
| Haberman | 3 | 2 | 225,81 | 306 |
| Heart | 13 | 2 | 150,120 | 270 |
| Glass | 9 | 6 | 70,17,76,13,9,29 | 214 |

Although benefits of NS theory and some aspects of data clustering such as certainty is considered in the proposed cost function, cost function minimization suffers from local optimum points. Finally, although the proposed certainty definition in neutrosophic domain is appropriate for density-based and center-based clustering, it is not working well on non-density cases such as contiguous-based and nonlinear-shape clusters.

Table 3. Clustering accuracy for FCM, PCM, PFCM, HPFCM and Proposed method with 4 datasets: Ionosphere, Haberman, Heart and Glass.

| Methods | Ionosphere | Haberman | Heart | Glass |
|---------|-----------|----------|-------|-------|
| FCM | 70.94 | 51.96 | 51.31 | 42.08 |
| WEFCM | 76.58 | 77.12 | 72.88 | 54.39 |
| PFCM | 70.94 | 51.96 | 51.31 | 42.08 |
| Soft-DKM | 67.77 | 51.42 | 50.24 | 40.50 |
| CDFCM | 75.26 | 74.68 | 71.95 | 52.96 |
| DPFCM | 75.26 | 76.50 | 71.89 | 53.33 |
| **Proposed Method** | **79.12** | **80.04** | **83.34** | **59.43** |

## VI. CONCLUSION

In this research, an effective clustering method was proposed in NS domain. For this task, data certainty was proposed based on density properties of data in NS domain to control outlier and boundary points followed by proposing a cost function in NS domain. Two types of clusters including main clusters and noise cluster are considered in the proposed cost function. Experiments on 7 different datasets including Scatter and UCI datasets showed that the proposed method not only handles outlier and boundary points but also outperforms existing clustering methods. Future efforts will be directed towards introducing certainty in NS domain for supervised methods such as deep convolutional neural networks(CNNs). Future efforts will be also directed towards proposing methods in neutrosophic domain for handling contiguousbased and nonlinear-shape clusters.